# Determining van der Waals materials' optical and polaritonic properties using cryogenic FTIR micro-spectroscopy.

Siddharth Nandanwar[1,2], Aditya Desai[1], S. Maryam Vaghefi Esfidani[1], Tristan McMillan[1], Eli Janzen[3], James H. Edgar[3], and Thomas G. Folland[1*]

[1] Department of Physics and Astronomy, The University of Iowa, Iowa City, Iowa, 52245

[2] Department of Physics, Boston College, Chestnut Hill, Massachusetts 02467, USA

[3]Tim Taylor Department of Chemical Engineering, Durland Hall, Kansas State University, Manhattan, KS 66506

*thomas-folland@uiowa.edu

## Abstract

Van-der-Waals materials have been shown to support numerous exotic polaritonic phenomena originating from their layered structures and associated vibrational and electronic properties. This includes emergent polaritonic phenomena, including hyperbolicity and exciton-polariton formation. However, many van-der-Waals materials' unique properties are most prominent at cryogenic temperatures. This presents a particular challenge for polaritonics research, as reliable optical constant data is required for understanding light-matter coupling. For infrared polaritonics (3-100μm), the small size of exfoliated flakes makes conventional ellipsometry impossible. This paper presents a cryogenic Fourier transform infrared microscope design constructed entirely from off-the-shelf components and fitting procedures for determining optical constants. We use this microscope to present the first temperature-dependent characterization of the optical properties of hexagonal boron nitride grown with isotopically pure boron. We show that Fabry Perot-type resonances close to the transverse optical phonon show the key temperature-dependent tuning of several parameters. Our full analysis of the infrared dielectric function shows small but significant tuning of the optical constants, which is highly consistent with Raman data from the literature. We then use this dielectric data to perform and analyze the polariton propagation properties, which agree extremely well with published cryogenic scattering-type nearfield microscopy results. In addition to the insights gained into hyperbolic polaritons in hBN, our paper represents a transferable framework for characterizing exfoliated infrared polaritonic materials and other infrared devices. This could accelerate discoveries in other material systems, especially those that are spatially inhomogeneous or cannot be prepared as large single crystals.

**Keywords;**

Phonon polaritons; Boron Nitride; Optical Properties; Cryogenic Spectroscopy; Vibrational Spectroscopy

**Introduction**

In recent years, research into mid- and far-infrared (3-100μm) polaritonics has rapidly expanded owing to the emergent physical phenomena in this wavelength range. Notably, this spectral range includes information on lattice vibrations, free carriers, and interband transitions in these materials[1]. One growth area has been the study of van-der-Waals (vdW) materials with a naturally layered structure[2,3]. This layered structure has several fascinating physical effects, including hyperbolicity[4–7], ballistic transport[8], and other exotic quantum phenomena. Furthermore, they can be prepared from bulk crystal samples through mechanical exfoliation, producing high-quality thin films on arbitrary substrates[9]. While many of the exciting properties of vdW materials can be accessed at room temperature, numerous effects, such as phase transitions[10], high mobility behavior[8,11], carrier freeze-out[12], and damping pathways[13], cannot be studied at room temperature. These effects are essential for emerging quantum technologies[14], in which polaritonics play a crucial role[15]. Therefore, developing techniques for studying infrared polaritonics in vdW materials at low temperatures becomes a key challenge. This is not trivial, as the wavelength of light is commensurate with the size of the crystals prepared. Any system capable of performing these studies must have sufficient spatial resolution and the capability to work at reduced temperatures. This work presents the design rules for a Fourier Transform Infrared cryo-microscope, which can perform spectroscopy over the mid-infrared at low temperatures and spatial resolutions commensurate with 2D flakes. We use this microscope to measure the temperature infrared dielectric function of hexagonal boron nitride (hBN) enriched with pure $^{10}$B and $^{11}$B and the role of polariton propagation. Our approach does not require specialized optical components or cryostats, which will significantly expand opportunities in low-temperature micro-spectroscopy of materials.

Existing low-temperature infrared optical characterization techniques can be split into near–and far–field categories. Near-field techniques typically exploit a sharp metal probe to get a precise spatial resolution (10nm) determined by the probe at a wide range of wavelengths from the infrared to THz[16]. They can operate at low temperatures and have broad capabilities in determining the properties of polaritons in van der Waals materials[8]. However, the reliance on laser sources for specific wavelength ranges and the possibility of artifacts in the data[17], make it less versatile for general-purpose characterization or materials screening. Furthermore, quantitative polarimetry is currently challenging, as tip scattering naturally favors certain field components preferentially[18]. Far-field techniques are diffraction-limited and, therefore, cannot probe materials with lateral dimensions below the wavelength of light in free space. However, they make up for what they lose in spatial resolution in bandwidth and polarization control and can provide ultra-broadband information in a single spectrum spanning the entire mid-infrared[1,19]. They are also better suited to larger samples where hyperspectral imaging is required. One approach is cryogenic infrared ellipsometry, but this typically requires samples on the order of one square cm to function[20]. FTIR microscopy loses some precision in polarization but can deal with individual van-der-Waals flakes[21]. Several approaches to FTIR micro cryo-spectroscopy for small samples have been developed in past works[11,22–28]. One of the most successful techniques has been parabolic, focusing on sub-mm crystallites and subsequent gold overcoating for referencing[29]. It has been widely used in numerous papers and gives accurate information from relatively large crystals[23,30,31]. However, it is unsuited for single, exfoliated flake analysis at a spatial resolution of 100μm or less. Transmission-based focusing setups leveraging light pipes are also effective on this length scale[27,28]. However, these cannot perform reflection-based spectroscopy, typically do not define their spot size well, and are not suited to mapping. Finally, several groups have adapted a commercial He cryostat to a commercial FTIR microscope and have been used to study van-der Waals materials[11,22,25]. However, this approach needs more flexibility to reconfigure optical systems to improve throughput and perform accurate polarimetry and requires expensive commercial microscopes.

This paper presents the design and principles for an infrared cryo-microscope that leverages off-the-shelf components capable of performing precision low-temperature and dielectric function measurements. Our approach can be applied to any combination of FTIR and optical access cryostat and optimized for polarimetry and other experiments. As a proof of concept for the study, we examine the dielectric function of $^{10}$B and $^{11}$B enriched hexagonal boron nitride. These materials have been shown to have record-breaking propagation lengths for hyperbolic phonon polaritons (HPhPs), owing to the reduced scattering in isotopically pure materials[13,17,32–34]. Both h$^{10}$BN and h$^{11}$BN have extended polariton lifetime at low temperatures at a given frequency[13,35]; however, the full dielectric function was not extracted. While such information is available for hBN with naturally distributed boron isotopes[22], the damping parameters, differences, and phonon tuning are key for extracting polariton lifetimes and changes in dispersion. We examine two substrate choices for evaluating the dielectric function of exfoliated flakes: Au mirrors and sapphire. The former has a strong background reflectance from the substrate aids in referencing. It provides well-defined Fabry-Perot (FP) type absorption resonances[36,37] in the mid-IR, which is well suited to determining the temperature dependence of TO phonons in the infrared. The latter is better suited for completing a full dielectric function analysis close to both TO and LO phonon energies. We perform a dielectric function analysis, extending previous methods[7,21] and implementing a method for backside reflection correction. We also address the sensitivity of the fitting procedure to specific parameters in the dielectric function and find that our method gives results consistent with past measurements on hBN polaritons[35] and Raman modes[13]. Our results for the full dielectric function show comparable shifts in the TO phonon energy as the frequency shift of the FP resonances, suggesting that tracking the properties of FP modes is a suitable mechanism for estimating the temperature dependence of some dielectric function parameters. Finally, we calculate the quality factors of polaritons as a function of temperature and show that our temperature-dependent dielectric function closely matches prior experiments by Ni[35] This suggests our dielectric function improves on those previously published, even at room temperature. We also show that the temperature-dependent properties of HPhP modes arise purely from the properties of the phonons, not from the polariton-dependent scattering mechanism. Crucially, our approach to cryo-spectroscopy, using off-the-shelf components, can be adapted to any commercial FTIR with an external port. Our accurate dielectric information for isotopically pure boron nitride also provides insight into the properties of van-der-Waals materials at low temperatures.

**Results**

Our microscope implements a conventional, finite conjugate length optical microscope implemented using all-reflective optical components, with a schematic shown in **Fig. 1a**. Indicative full beam power spectra are shown in **Fig 1b**, comparing the spectra from the microscope with a 3mm aperture (measured 60μm spot size) to a conventional configuration FTIR (SiC Glow Bar, KBr beam-splitter, and DLaTGS detector). For more details, please see the methods section. We examine the low-temperature dielectric response of h$^{10}$BN and to demonstrate the cryo-microscope's capabilities for determining polaritons' properties at low temperatures. Temperature-dependent micro-FTIR spectra of a flake of boron h$^{10}$BN, which is 1220nm thick, exfoliated onto an Au-coated silicon substrates substrate at 300K and 5K are shown in **Fig. 2a.** There are resonances are observed in the spectral ranges of approximately 840cm$^{-1}$ and the range from 1200-1400cm$^{-1}$. The narrow spikes in the range of 1400-1800cm$^{-1}$ can be attributed to atmospheric water absorption – and are not considered for analysis. All resonances are purely absorptive, as the Au mirror does not transmit light in this spectral range. We attribute these to a Fabry Perot-type resonance associated with the high index of the h$^{10}$BN close to the TO phonon (located at 1393cm$^{-1}$), which is reported to exceed 25 in prior works[32]. These have been reported previously in the context of FTIR-ATR measurements[36], on bulk[38], and on exfoliated flakes[37,34]. Note that the out-of-plane resonances can be measured due to the reflective objective's off-normal incidence angle. There are no significant changes in the mid-infrared spectrum between 5K and room temperature, as anticipated based on prior studies on h$^{10}$BN[13,35], but minor frequency and line width changes are associated with the phonons.

A detailed view of the resonance associated with the out-of-plane phonon and the lower Reststrahlen band of h$^{10}$BN is shown in **Fig. 2b**. The linewidth is exceptionally sharp, consistent with the previously reported long lifetime of the out-of-plane phonon[32]. This resonance has a minimal spectral shift with temperature, suggesting a relatively weak dependence of the out-of-plane phonon modes on temperature. A detailed view of resonances associated with the high energy phonon is presented in **Fig. 2c**. In this range, a series of dips in the spectra are present, which get increasingly sharp as they approach the TO phonon located at approximately 1393cm$^{-1}$ before merging into a broad peak with relatively weak absorption. First, we consider the resonance frequency of these modes - they redshift with increasing temperature, with a tuning range of 1.8±0.1cm$^{-1}$, 2.7 cm$^{-1}$±0.1and 2.5±0.1cm$^{-1}$, for modes FP1, FP2, and FP3 respectively. This is suggestive of temperature tuning of the TO phonon in hBN. The trend in linewidth between the modes is more subtle. The linewidth of standing wave resonances is directly related to a combination of material and radiative damping. As the refractive index increases close to the TO phonon, radiative damping decreases due to increased reflection at the interface between hBN and air. Meanwhile, light more readily gets absorbed by the TO phonon, so material absorption rises dramatically. This results in resonances getting sharper as they approach the TO phonon as the radiative losses decrease but then get significantly broader when the TO phonon absorption increases. There is a small but notable change in the model linewidth with temperature - taking FP3 as an example, we can see that the full-width half maximum changes from 3.2 cm$^{-1}$ to 3.7 cm$^{-1}$ between 5K and 300K. This is an indication that there is a small but notable change in the mode lifetime.

These results are compared with temperature-dependent micro-FTIR spectra of a flake of boron hB$^{10}$N with 700nm thickness exfoliated onto a sapphire substrate, as shown in **Fig. 3a.** We see several spectral features across the mid-infrared, including the phonon response of the sapphire substrate (900cm$^{-1}$ and below), a sharp dip associated with the out of plane phonon of hBN at approximately 842cm$^{-1}$, and the in-plane Reststrahlen band from approximately 1395cm$^{-1}$ to 1650cm$^{-1}$, and sapphire below 900cm$^{-1}$. The resonance associated with the out-of-plane phonon of h$^{10}$BN can be seen in the Reststrahlen band of sapphire, as shown in **Fig 3b**. Due to the high reflectivity of sapphire in this region, the results are very similar to those present on an Au substrate and match our previous results well. In the hBN upper Reststrahlen band, **Fig 3c**. Each mode redshifts with increasing temperature, with the mode labeled FP1 tuning 2.5±0.5cm$^{-1}$ and FP2 tuning 2.6±0.1cm$^{-1}$. These can be directly contrasted with the results for FP3 on Au, which shows 2.5±0.1cm$^{-1}$ of tuning. Similarly, the linewidth slightly broadens with temperature, going from 6.8±0.3cm$^{-1}$ to 7.3±0.3cm$^{-1}$. This consistency between substrates suggests a phonon tuning of approximately 2.5cm$^{-1}$, and a reduction of the phonon lifetime. Similar results are seen for a flake of isotopically pure h$^{11}$BN and a second thinner flake of h$^{10}$BN, detailed in **Supplemental**

Section 1. However, directly proving the temperature dependence requires extracting the infrared dielectric function, which can be performed on sapphire's upper Reststrahlen band.

**Discussion**

To determine the optical constants of the hBN, we use both a commercial piece of software WAVSE from J. A. Wollam, and homebuilt. a nonlinear least squares fitting algorithm based on a 4x4 transfer matrix formalism published in [39], with both giving consistent results. Both vary the dielectric function parameters and construct transfer matrices that match the best fit to the reflectance data. The incident angle (20º), and polarization are set parameters, and the substrate uses room temperature literature values from [40]. The thickness of the flakes is determined by AFM, but is allowed to be fitted in a 5% range about the AFM measured values. The optimization first constructs the projected dielectric function for the material of interest using starting input parameters. The in-plane and out-of-plane dielectric function for hBN is modeled using the form:

$$\varepsilon = \varepsilon_\infty \frac{\omega_{LO}^2 - \omega^2 - i\Gamma\omega}{\omega_{TO}^2 - \omega^2 - i\Gamma\omega} \quad (1)$$

Where $\varepsilon_\infty$ is the high-frequency permittivity, $\omega_{TO}$ is the transverse optical phonon, $\omega_{LO}$ is the longitudinal optical phonon, and $\Gamma$ is the phonon damping. Both in-plane and out-of-plane dielectric function parameters use the form above, with out-of-plane values fixed to the values from Giles[32], and in-plane values fitted. The optimization then constructs the system's transfer matrices using this dielectric function and the fixed parameters for the rest of the system. The projected reflectance across the wave numbers from the experiment file is calculated and compared to the experimental reflectance using the least squares method. The above dielectric function parameters are then optimized iteratively. Error bars are constructed from the Jacobian of the fit, and the experimental and projected reflectance spectra are plotted to see the precision of the optimization. Only the in-plane dielectric function is fitted, as we can only resolve spectral features from the full Reststrahlen band in the in-plane direction. We used published data for the out-of-plane dielectric function[32]. The fitting includes three experimental non-idealities: angular spread, reflection scaling, and back-reflection compensation. Angular spread is achieved by averaging several spectra across the incident beam (10º). A scaling factor is included to account for the non-planarity of the sample with the microscope objective. Back reflection from the cold finger underneath the flake interfered with the spectra at higher wave numbers, as discussed in **Supplementary Section 2**. The influence of this back reflection was subtracted from the spectrum before fitting. The fit is constrained to 1000 to 2000cm$^{-1}$, reducing the area that incorporates back reflections. Finally, as the measurement is most accurate where the light-matter interaction is strong about the TO phonon, this area is prioritized in the fit through error bars of 0.01 in an 80cm$^{-1}$ bandwidth around the TO phonon value and 0.05 outside this value.

Spectral fits for the reflectance spectrum at 5K, including a broader frequency range outside the fit range centered around the TO phonon, are shown in **Fig 4a/b**. There is an excellent agreement between the modeled results and experimentally measured reflectance spectra close and below the TO phonon, with some error occurring above the Reststrahlen band. This is attributed to reflections off the backside of the substrate, which increases the reflectance signal and motivates our choice of a narrow region for fitting. We conducted full dielectric function analysis as a function of temperature for two h$^{10}$BN flakes and one h$^{11}$BN enriched flake, fitting spectra for temperatures between 5.5 and 300K. Fits are shown in **Supplemental Section 3.** The extracted temperature-dependent parameters for $\omega_{TO}$ and $\Gamma$ are shown in **Fig 4c-d,** distinguishing between the two isotopes used. The phonon energy continuously redshifts with increasing temperature for all flakes studied here, consistent with prior Raman studies by Cuscó. There is a small inconsistency between the two flakes presented here, likely due to systematic errors associated with the fitting process and the resolution of the FTIR at 0.5cm$^{-1}$. Through the deviations between the two parameters, we can estimate systematic errors at 0.9cm$^{-1}$. We highlight that our results show a slightly lower room temperature phonon energy than that previously reported in the literature by Giles[32]. This is likely attributed to the precision of the measurements in Giles [32] at 2cm$^{-1}$ – ours are reported at a higher spectral resolution of 0.5cm$^{-1}$, presenting more accurate values than those for the previous work. This demonstrates that our cryogenic measurements and fitting approach for TO phonons are accurate within the spectral resolution of the FTIR. We can compare the total temperature tuning of the FP modes against those extracted from dielectric data, which shows that the TO phonon tuning can be inferred from the shifting of the TO phonon modes.

Our results for damping $\Gamma$ are shown in **Fig 4d-e** are also consistent with recent Raman and s-SNOM studies, showing that the phonon linewidths are reduced as the samples are cooled. Our results for damping are slightly larger than those reported in Giles[32]. We can attribute to the use of a harmonic model in Giles \[32] for the analysis of bulk-like flakes where an anharmonic model is more applicable[24]. Again, we see some inconsistency between the fits for the two B[10] flakes, which we can use to estimate systematic errors during the fit of 0.47cm$^{-1}$. To better compare our results with those from Ni[35], we use the temperature-dependent fitting function proposed in that work;

$$\gamma(T) = \gamma_0 \left(1 + \frac{8\pi^2}{15} \frac{T^4}{T_\lambda^4}\right) \qquad (2)$$

Where T is the temperature, $\gamma_0$ is the low-temperature damping limit, and $T_\lambda$ is the characteristic temperature. Fitting to the data in Figure 4, we get a characteristic temperature of 1200±100K for the 700nm h$^{10}$BN flake, 1110±60K for the 500nm thick h$^{10}$BN flake and 1060±30K for the 750nm thick h$^{11}$BN flake. These values are highly consistent and support the hypothesis that the damping is largely given by acoustic phonon scattering. While this value is approximately twice the value presented by Ni[35], this is potentially due to uncertainties in other parameters in their model. This will be explored later in the paper by explicitly calculating the temperature dependence of the phonon polaritons. Finally, the change in the line widths of the FP modes in the reflectance spectrum on Au or Sapphire gives a comparable change to those extracted from the dielectric function analysis. As such, we believe that fitting the line shape of FP modes close to the TO phonon can approximate the temperature-dependent properties of $\omega_{TO}$ and $\Gamma$.

While our results for $\omega_{TO}$ and $\Gamma$ are in excellent agreement with prior Raman literature and extend the previously reported values from infrared studies, we find that both $\omega_{LO}$ and $\varepsilon_\infty$ parameters show some small inconsistencies (**See Supplementary Section 4**). While close to literature-reported values, suggesting our fit is physical, we do not have the accuracy in resolving the temperature tuning. This is attributable to the back reflection from the substrate's bottom and the relatively thin nature of the flakes exfoliated in this work. Our backside correction cannot fully account for this error, but it does lead to more physical results consistent with prior studies on hBN than without. We estimate the systematic error on $\omega_{LO}$ to be of the order of 3cm$^{-1}$ based on the standard deviation of the data, insufficient to resolve the few wavenumbers shift we would anticipate for phonon tuning for hBN going to cryogenic temperatures. Similarly, we estimate the systematic error on $\varepsilon_\infty$ being of the order of 0.1. While this is a limitation of our approach, we show in **Supplementary Section 5** that our dielectric function can reproduce the FP mode tuning from the flake in Figure 2, suggesting that our model is accurate enough for predictive modeling.

Finally, we assess the properties of the hyperbolic phonon polaritons in hexagonal boron nitride. To do so, we will evaluate the dispersion of the modes, mode quality factors, and damping rates. The expression gives the dispersion of hyperbolic polaritons in the high k limit[4];

$$kd = (k' + ik'')d = -\psi \left[ \operatorname{atan}\left(\frac{\varepsilon_o}{\varepsilon_t \psi}\right) + \operatorname{atan}\left(\frac{\varepsilon_s}{\varepsilon_t \psi}\right) + \pi l \right], \quad \psi = -i\sqrt{\frac{\varepsilon_z}{\varepsilon_t}} \quad (3)$$

Where d is the flake thickness, $\varepsilon_o$ is the dielectric function above the hBN flake, $\varepsilon_s$ is the dielectric function of the substrate, $\varepsilon_t$ is the in-plane dielectric function of hBN, $\varepsilon_z$ is the out-of-plane dielectric function, and $l$ is an integer, set to 1 here. We can first examine the polariton dispersion - given by the real part of k, and the mode Q factor (k'/k'') as shown in **Fig. 5a/b.** To simplify the presentation of the data, we only consider the extreme cases of 300K and 5K, and calculate for both suspended hBN and on SiO$_2$. Suspended hBN has only the intrinsic properties of the flake, while SiO$_2$ is often used as a substrate for van der Waals materials. We observe that the dispersion does change slightly between room temperature and 5K, but not significantly. Notably, the choice of suspended versus SiO$_2$ substrates contributes more to the dispersion than the temperature shift. However, the mode quality factors show a dramatic shift in behavior, both with substrate and with temperature. The presence of absorption in the SiO$_2$ substrate and a change in the group velocity degrades the mode quality factor, as discussed previously[41]. The reduction in the losses associated with the phonon when the sample is cooled produces a quality factor approximately 30% larger than at room temperature in the center of the Reststrahlen band. This enhancement is reduced on SiO$_2$ substrates, suggesting that at these frequencies, the losses in SiO$_2$ start to become more significant and suppress the benefits of cryogenic cooling. While this is not a dramatic change in mode quality factor, this is attributable to the only accessible loss factors being acoustic phonon scattering by the TO phonon, as previously reported. We can compare our results against quality factor values taken from Ni[35], and we show extremely good agreement between our modeled values and experiment. These represent the closest agreement between s-SNOM and far-field studies reported for isotopically pure hBN, suggesting that our higher damping values than those of Giles[32] represent the true loss in isotopically pure hBN.

To further explore the correlation between far-field and near-field measurements, we also calculate the polariton damping factor, which can be evaluated from[35] $\gamma(T) = \omega v_g/(Q(T)v_p)$, where $v_g$ and $v_p$ are the group and phase velocities of the polariton, respectively. Both $v_p$ and $v_g$ are evaluated from the dispersion in Fig 5a. Our temperature-dependent polariton data is shown in **Fig 5c**. Our results for suspended h$^{11}$BN are consistent with the published data, which corrects for the damping of the substrate. We can also evaluate the substrate-dependent damping, which is shown to be 0.2cm$^{-1}$ at a frequency of 1522cm$^{-1}$. We can compare the temperature dependence of the polariton damping against the temperature dependence of the phonon damping by fitting to the same functional form. We find a characteristic temperature $T_\lambda = 1060 \pm 30$K identical to the phonon value. This suggests that the previously reported damping associated with hyperbolic phonon polaritons in boron nitride is almost entirely given by losses to the phonon part of the polariton. Based on polariton imaging alone, this is impossible to determine, and it demonstrates the complementary nature of near- and far-field measurements.

## Conclusions

In this paper, we have detailed how to design a cryogenic FTIR microscope capable of measuring the infrared optical properties of exfoliated van-der-Waals materials. The instrument is constructed using commercial FTIR and off-the-shelf components, which can be implemented at a fraction of the cost of commercial systems. We leverage this instrument to measure the temperature-dependent properties of $^{10}$B and $^{11}$B isotopically pure hBN crystals. We find spectral features associated with FP resonances, such as spectral tuning and sharpening, attributed to the tuning of the TO phonon and a reduction in phonon linewidth. We performed fitting to evaluate the full infrared dielectric function as a function of temperature and show that our procedure gives excellent accuracy for TO phonon energies and damping. Our results are consistent with past Raman and polaritonic studies and suggest that the polariton losses at room temperature have been underestimated in prior works. By analyzing the fitted dielectric function, we agree with the experimentally measured properties of HPhPs at low temperatures. This suggests that the changes in mode propagation measured experimentally are associated with changes in the properties of the transverse optical phonon. Our work and methods can be applied beyond hBN to a much more comprehensive set of van der Waals materials, including those with phase transitions, free electron plasmas, and other optical properties. It can also be applied to various microelectronic devices with features on the tens of microns scale, which would otherwise be difficult to probe.

**Methods**

Our home built FTIR microscope uses a commercial FTIR spectrometer (Bruker Vertex 80V) with an incandescent source and an external output, which is then coupled to the microscope. Infrared light emitted by an incandescent source (SiC Glow Bar) illuminates a variable diameter pinhole (a Jacquinot Stop or J-Stop), typically set to 3mm for measurements in our system. Light passing through this pinhole is collimated by a parabolic mirror of focal length $f_{col}$ =15cm to a beam approximately 5cm wide and passed through an interferometer and a series of mirrors into the microscope setup (not shown in Fig. 1a). A flip mirror is included at the start of the microscope to allow white light illumination for sample identification. All mirrors used in the microscope are silver to allow operation into the visible for sample identification while maintaining good mid-infrared reflection. Light from the FTIR is focused by a long focal length concave mirror (500mm), forming an image of the J-Stop at the focus, which is aligned co-inside with the back-focal length of the finite conjugate objective. This image of the J-Stop is demagnified by the ratio between the focal length of the internal collimating mirror ($f_{col}$) and the concave mirror ($f_{con}$); in our case, this gives a nominal demagnification factor of x3.3 (from $f_{con}/f_{col}$) and a spot approximately 10mm across at the back focal plane. We note that the aberrations inherent to using a concave mirror off-axis are relatively minimal and do not significantly influence performance. At this position, we also place a pinhole with a 0-12mm calibrated aperture to define the illumination spot on the target.

Light then travels past a D-shaped half mirror, into a reflective objective (25x – Thorlabs LMM25XF-P01), and onto the sample. The illumination spot defined by the pinhole is magnified onto the sample, which is placed inside an Oxford Instruments LHe cryostat equipped with a Cold Edge Hydra closed cycle chiller with ZnSe or KRS5 windows. The light illuminating the sample is collected by the objective and reflected by a half mirror at the back of the objective. A half-mirror ensures broadband operation and maintains a reasonable throughput and field of view (FOV), unlike conventional beamsplitter. A second motorized pinhole (ELL 15) is placed at the objective's back focal plane after the D-shaped half mirror to define the collection area analyzed by the detector. Using two pinholes reduces the presence of scattered light in the spectrum. Motorized polarizers (ELL 14) are positioned alongside both pinholes to analyze the reflected polarization state and are equipped with KRS5 (Thorlabs WP25H-K) or ZnSe (Thorlabs WP25H-Z) polarizers. A magnetic mount mirror switches between a visible camera and an infrared detector. In the imaging position, an achromatic doublet lens focuses light onto a visible camera with a CCD area of ½''. We maximize the FOV to allow for easy sample identification by using a 4f configuration, enabling us to image the pinhole completely from 1mm through 12mm aperture sizes. Chromatic aberration is present between visible and IR ranges from the ZnSe window, which is compensated for in measurements by a focal shift of the cryostat between optical and infrared measurements. For infrared detection, we use a combination of mirrors and a 1'' 90-degree off-axis parabolic mirror with a focal length ($f_{det}$) of 1'', focusing on an MCT detector (IR Associates FTIR-22-0.25). The off-axis parabolic is located 30cm after the back focal length of the objective. The detector is mounted on a modular base plate so that it can be swapped out with other detectors for spectroscopy.

It is worth discussing the choice of components used in this microscope configuration and how they can be adapted for different experiments. The parameters to optimize for an FTIR microscope are light throughput, spectral resolution, and spatial resolution. To control these, one chooses the properties of the J-Stop, concave mirror, objective, detector, and collection optics. The J-Stop should be selected to maintain sufficient spectral resolution at a given wavelength for the chosen applications. In our case, we decided to achieve a spectral resolution of 0.5cm$^{-1}$ across the full spectrum of the MCT, which is adequate for most materials' science applications. A larger J-stop can be used if a high resolution is not required, providing additional light into the microscope. A concave mirror is then chosen to fill the pinhole (accounting for the demagnification factor) and better match the detector's acceptance angle, characterized via the f-number of the components. The finite conjugate length objective lens f-number is typically low, and matching these can improve light throughput in the system (f=0.04 for the objective and f=0.1 for the concave mirror). The third choice is the objective selection, which should consider the sample sizes that need to be analyzed and the pinhole size range – here, 0-12mm. Here, we anticipate analyzing samples from 20μm-200μm, commensurate with the size of vdW flakes, choosing 50x (1mm-10mm pinhole). The final choice is the detector and collection optics. For HgCdTe detectors, typically smaller detectors offer dramatically higher responsivity due to reduced noise in small active area detectors[42]. However, smaller detectors also require a higher magnification factor to couple to the detector, which is challenging with off-the-shelf parabolic reflectors, which have a minimum focal length of around 25mm. We use a 25mm focal length off-axis parabolic, located approximately 300mm behind the pinhole, providing a magnification of roughly 11X between the iris and the detector. We choose to use a 250μm x 250μm HgCdTe detector with a 22μm cutoff, which offers a good compromise between detector sensitivity and bandwidth. Sensitivity can be improved by the choice of detectors –we have implemented Si, Ge, InSb, and pyroelectric detectors in addition to the MCT to cover the far-infrared and visible between different configurations.

Isotopically pure $B^{10}$ and $B^{11}$ enriched hBN samples were grown using the flux growth method described in the paper [34]. Crystals were then exfoliated onto c-plane sapphire substrates and Au-coated silicon, and sufficiently large (>50μm x 50μm) flakes were identified for spectroscopy using optical techniques and atomic force microscopy (see images **Figure 2 and 3** or **Supplementary Section 2)**. Sapphire substrates were chosen as they offer a low refractive index at the wavelength of the in-plane phonons, which provides a relatively flat background for dielectric function extraction. Relatively thick flakes are required for accurate dielectric function extraction, thicker than 500nm, to capture the transverse and longitudinal optical phonon energies sufficiently. Substrates with flakes were mounted onto a copper plate using silver paint, which was then screwed onto the cryostat cold head with indium foil acting as a thermal contact. A silicon diode temperature probe is mounted on the backside of the cryostat cold head to accurately monitor the sample's temperature. An Au mirror was placed next to the sample to provide an in-situ reference for all reflectance spectroscopy, and spectra were collected with a 0.5cm$^{-1}$ spectral resolution.

## Acknowledgments

SN, TM, and AD acknowledge funding through the University of Iowa Office for Undergraduate Research. TGF acknowledges funding from NSF Grant 2318049, 'Phonon Polariton Based Infrared Optoelectronics' and ONR Grant N00014-23-1-2616, 'Ultrafast, nano-optic and temperature-dependent infrared (IR) probes for wide bandgap semiconductor characterization'. SMVE acknowledges funding through NSF grant Grant 2318049, 'Phonon Polariton Based

Infrared Optoelectronics', as well as the University of Iowa Startup funding. ONR award N00014-22-1-2582 provided EJ and JHE support for hBN crystal growth

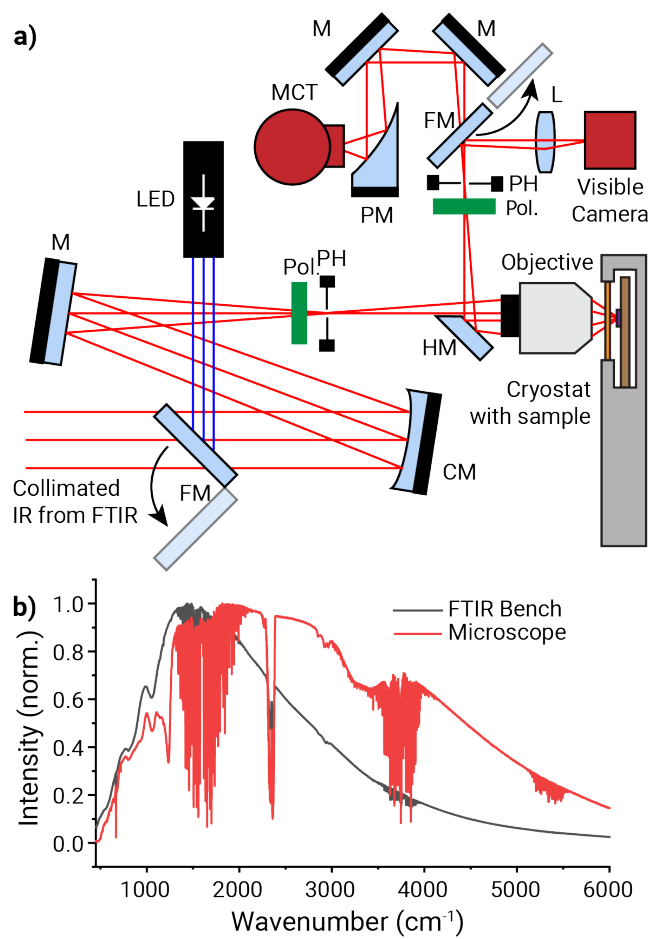

*Figure 1. Schematic of the FTIR cryo-microscopy setup. a) shows the optical configuration, with abbreviations M-mirror, CM – Concave mirror, PM – parabolic mirror, HM-Half mirror, FM – flip mirror, L- lens, LED – light emitting diode, Pol. – Polarizer and PH – Pinhole b) compares the comparative spectral throughput of our system against a DLaTGS detector in the bench of our FTIR to the throughput of the microscope.*

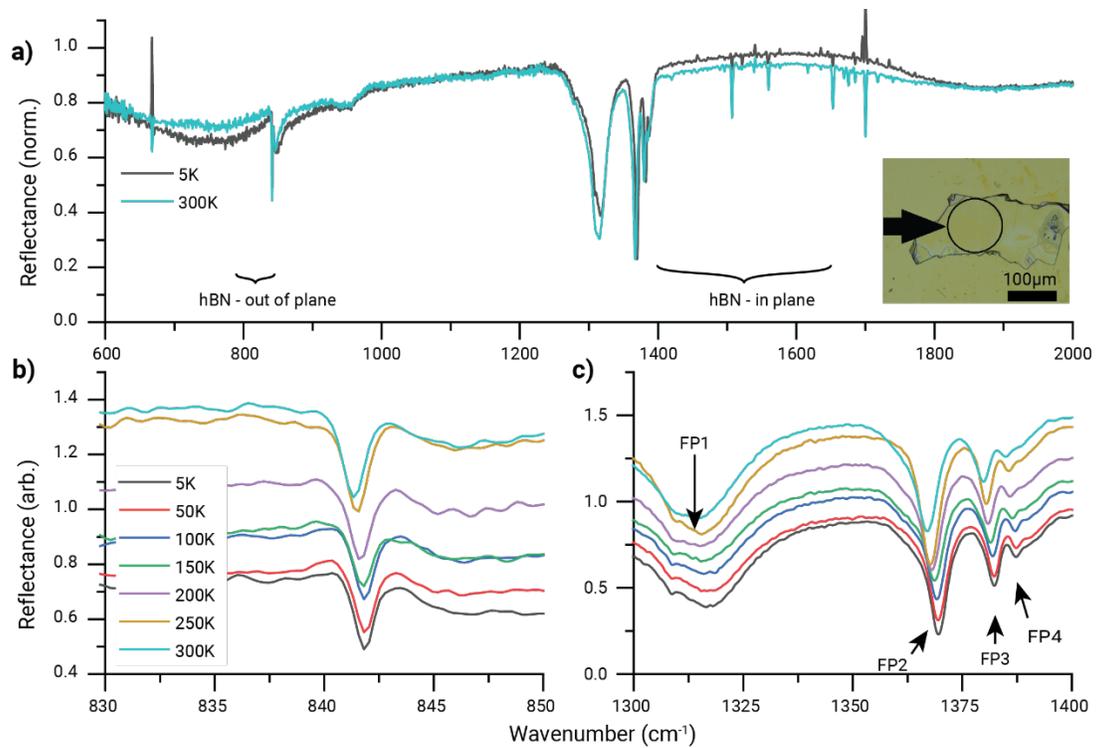

*Figure 2. Temperature-dependent infrared spectra of isotopically enriched boron nitride on Au/Ti coated silicon substrates. **a)** shows broadband reflectance spectra across the mid infrared, **b)** shows detail associated with the out-of-plane phonon in h$^{10}$BN **c)** shows detail associated with the in-plane phonon of h$^{10}$BN. Inset shows an optical microscope image of the flake.*

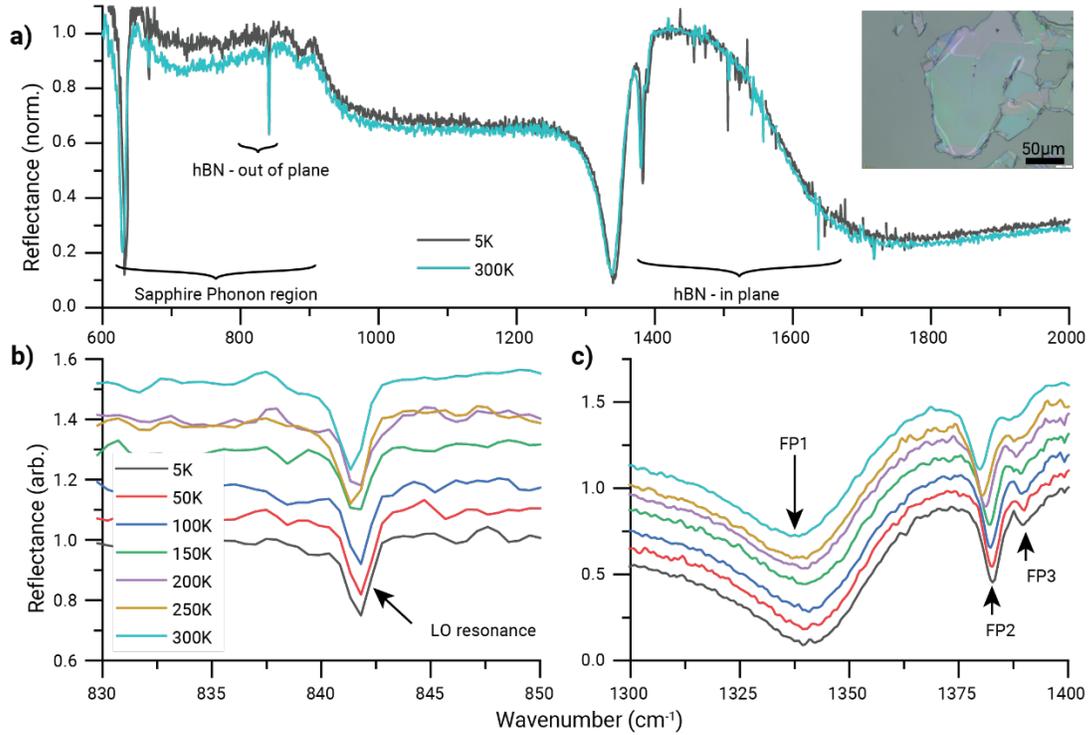

*Figure 3. Temperature-dependent infrared spectra of isotopically enriched boron nitride on Sapphire substrates. **a)** shows broadband reflectance spectra across the mid infrared, **b)** shows detail associated with the out-of-plane phonon in $h^{10}BN$ **c)** shows detail associated with the in-plane phonon of $h^{10}BN$. Inset shows an optical microscope image of the flake.*

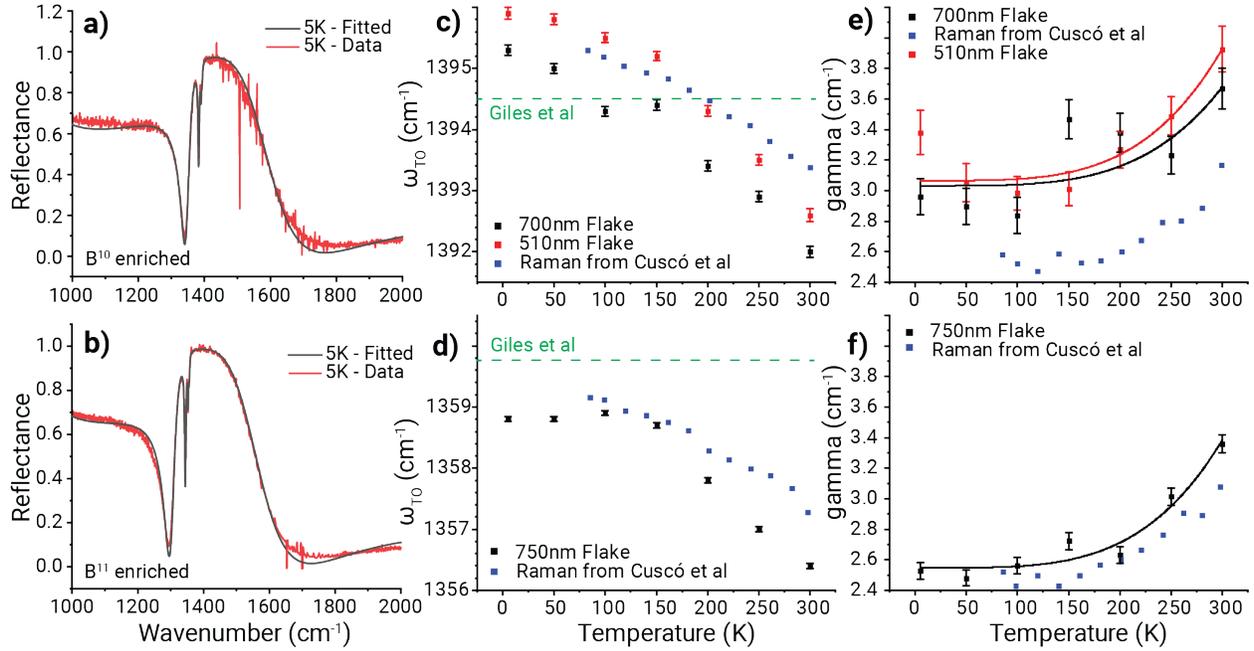

*Figure 4. Dielectric function extraction from hBN flake reflectance spectra **a)/b)** shows fitted spectrum for a 700nm thick h$^{10}$BN and h$^{11}$BN enriched flake respectively **c)/d)** shows experimentally extracted TO phonon values for h$^{10}$BN and h$^{11}$BN respectively **e)/f)** shows experimentally extracted phonon damping values h$^{10}$BN and h$^{11}$BN, respectively.*

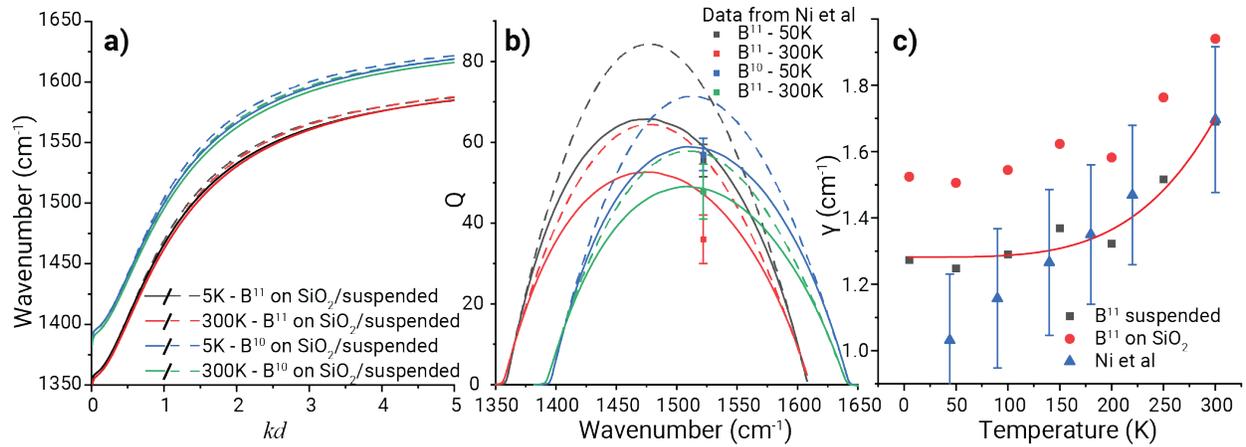

*Figure 5. Polariton dispersion and lifetime extracted from measured dielectric functions a) shows polariton dispersion for $h^{10}BN$ and $h^{11}BN$ on both a SiO2 substrate and suspended in air. b) shows the mode Q factor, defined as k'/k'', for $h^{10}BN$ and $h^{11}BN$ on both a SiO2 substrate and suspended, compared against data from Ni[35]. c) shows polariton damping rate and a temperature-dependent fit from our dielectric function, again compared against Ni[35] for $hB^{11}N$.*

# Supplemental information for 'Determining van der Waals materials' optical and polaritonic properties using cryogenic FTIR micro-spectroscopy.'

**Supplemental Section 1 – Reflectance Spectra for other Flakes studied.**

In this section we show full infrared reflectance spectra for a 510nm flake of $h^{10}BN$ and a 750nm thick flake of $h^{11}BN$.

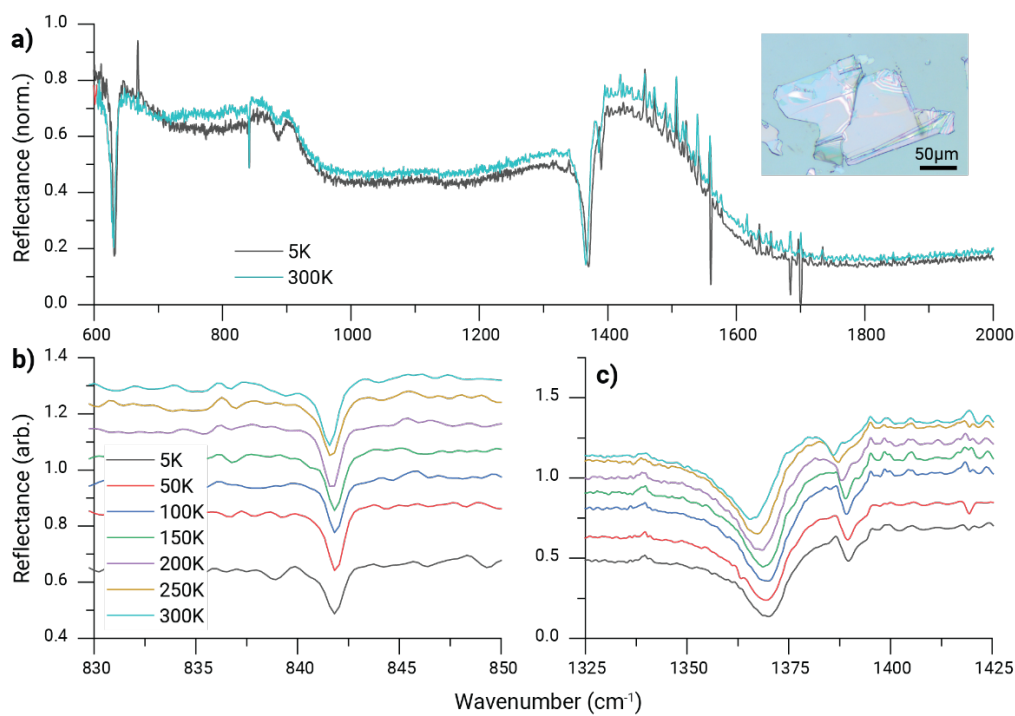

*Figure S1. Temperature-dependent infrared spectra of $h^{10}BN$ isotopically enriched boron nitride on Sapphire substrates. **a)** shows broadband reflectance spectra across the mid infrared, **b)** shows detail associated with the out-of-plane phonon in $h^{10}BN$ **c)** shows detail associated with the in-plane phonon of $h^{10}BN$. Inset shows an optical microscope image of the flake.*

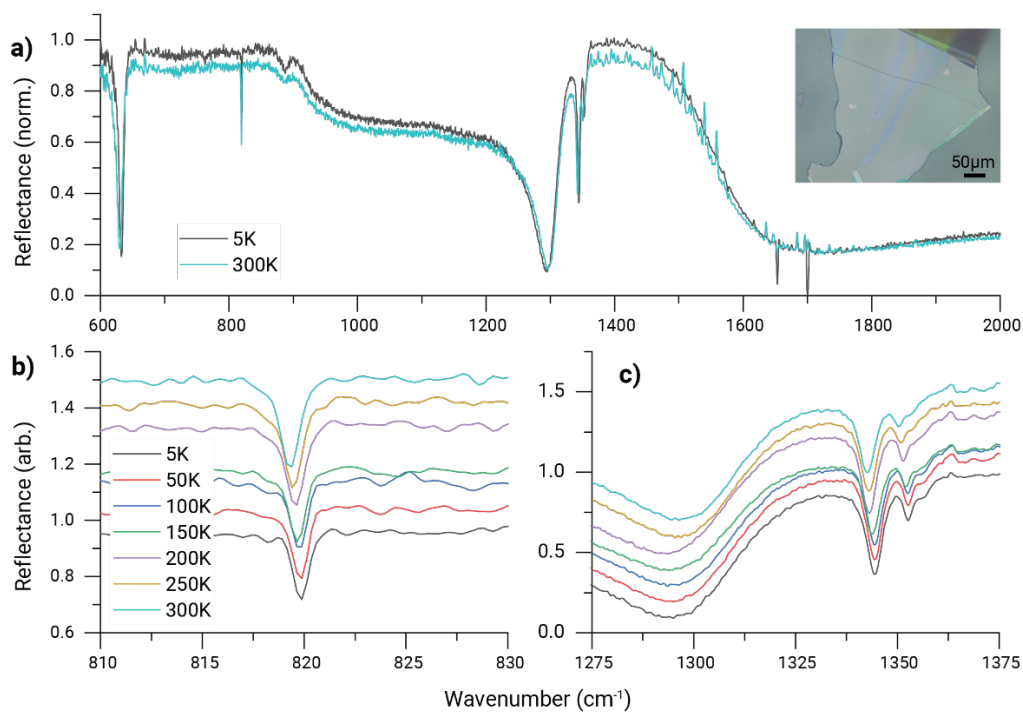

*Figure S2. Temperature-dependent infrared spectra of h$^{11}$BN isotopically enriched boron nitride on Sapphire substrates. **a)** shows broadband reflectance spectra across the mid infrared, **b)** shows detail associated with the out-of-plane phonon in h$^{11}$BN **c)** shows detail associated with the in-plane phonon of h$^{11}$BN. Inset shows an optical microscope image of the flake.*

**Supplemental Section 2 – Role of Back Reflections**

We have experimentally found that the copper plate on which the sample is mounted can contribute significant back reflections into the sample spectrum. Figure S3 shows the sapphire substrate's experimental and simulated infrared reflection. We see significantly higher observed reflection in the transmission window of sapphire. We can subtract this back reflection from the measured spectra by transmitting a 1mm sapphire substrate to correct this. To do this, we need to assess the strength of the back reflection, which is done empirically for consistency with prior published results. We find a value of ~20% gives physical values the reflectance at about 2500cm$^{-1}$ wavenumbers. This correction's influence is compared against simulated data in Figure S3, suggesting it is a good approximation for removing back reflections. We highlight that as we can fit the back reflections from the substrate data, this can be applied to any van-der-Waaals flake. This could also be corrected by techniques such as backside roughening or mounting the sample above holes in the cryostat.

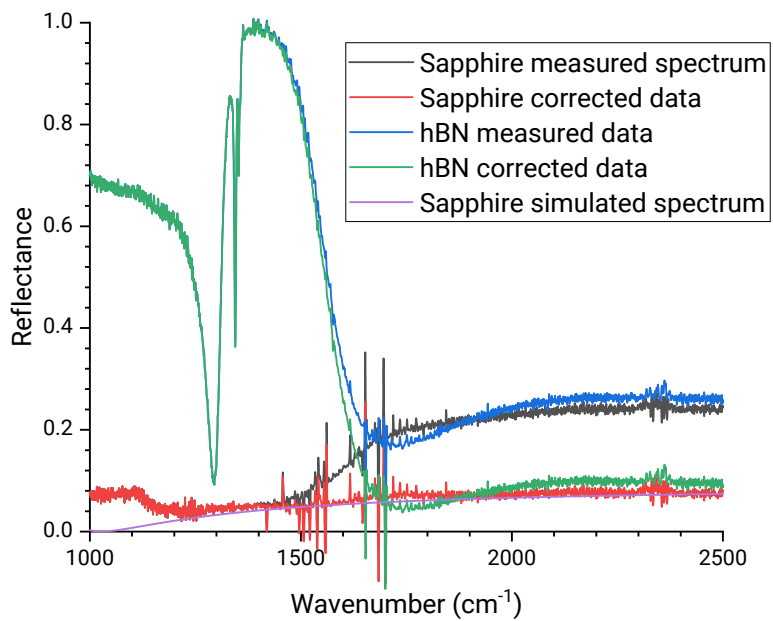

*Figure S3. Role of back reflections and correction factor introduced to the spectrum.*

**Supplemental Section 3 – Temperature Dependent Fit Data**

In this section, we present numerical fits for the temperature-dependent spectra over the fitted range for the three samples studied in this work. Figure S4 shows fit data for the 700nm h$^{10}$BN flake, Figure S5 for the 510nm h$^{10}$BN flake, and Figure S6 for the 750nm h$^{11}$BN flake. The thinner flake shows fewer FP modes than the thicker flake presented in the main text.

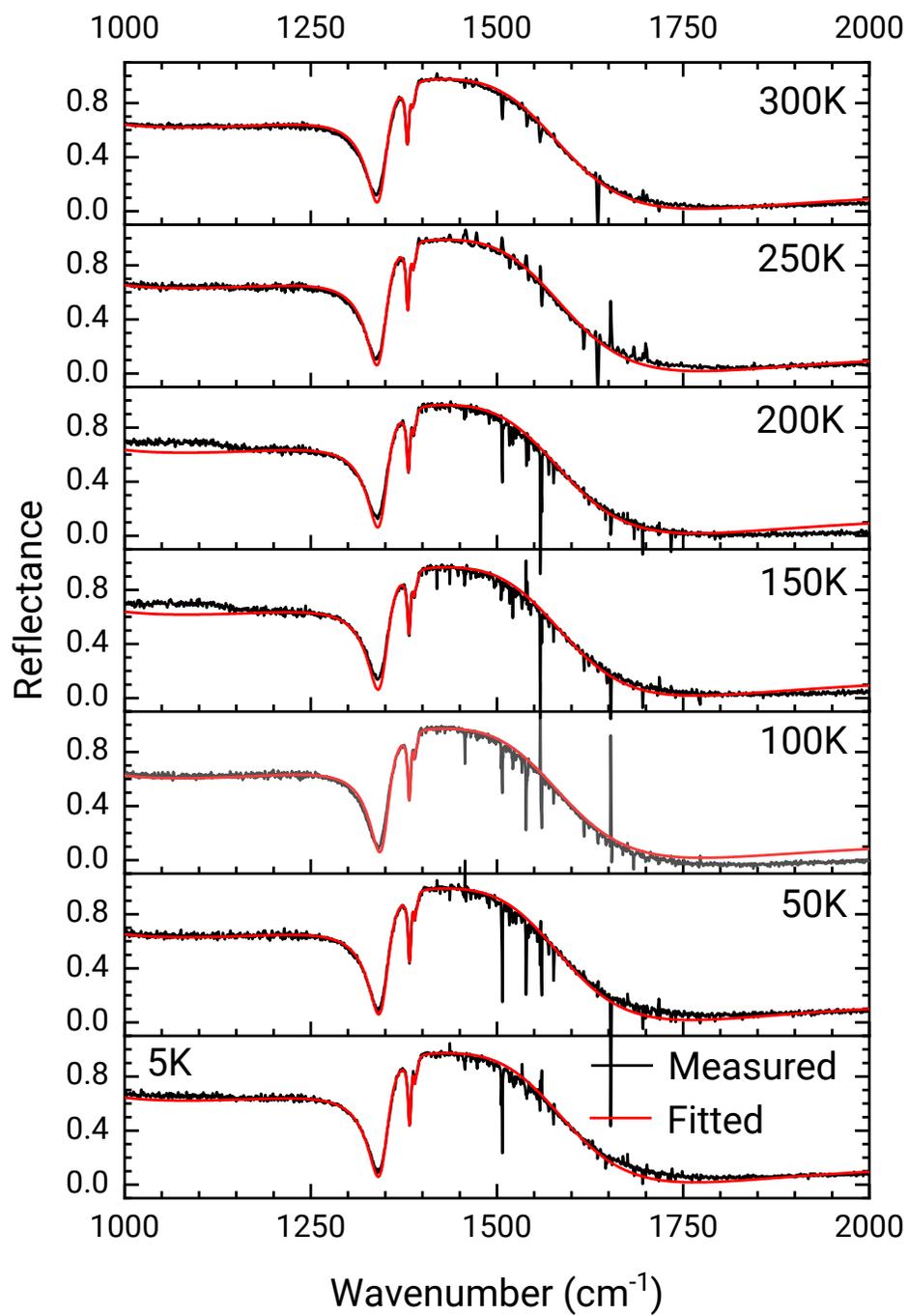

*Figure S4. Temperature-dependent spectra and fit data for the* 700nm h$^{10}$BN flake

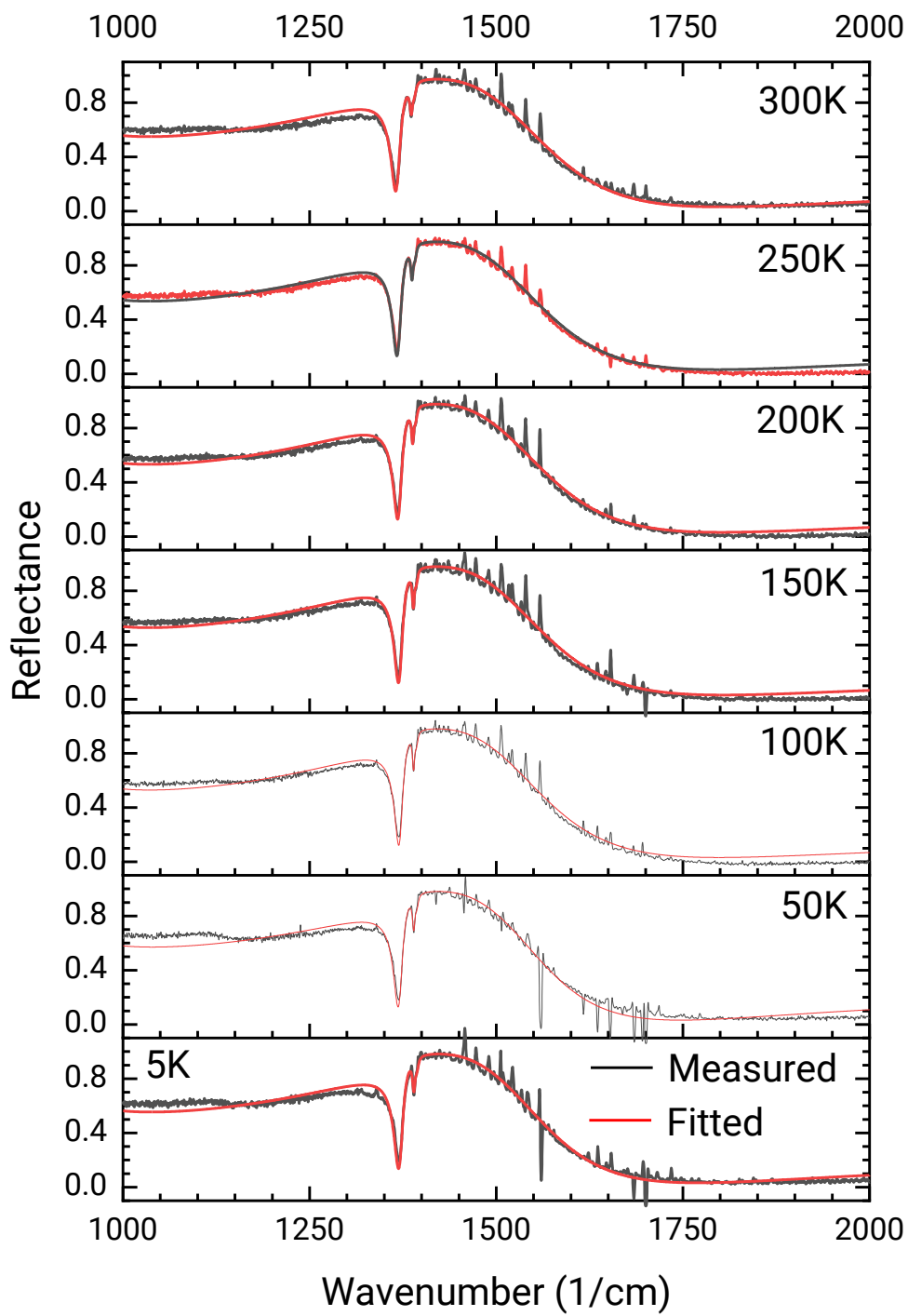

*Figure S6. Temperature-dependent spectra and fit data for the 500nm h$^{10}$BN flake*

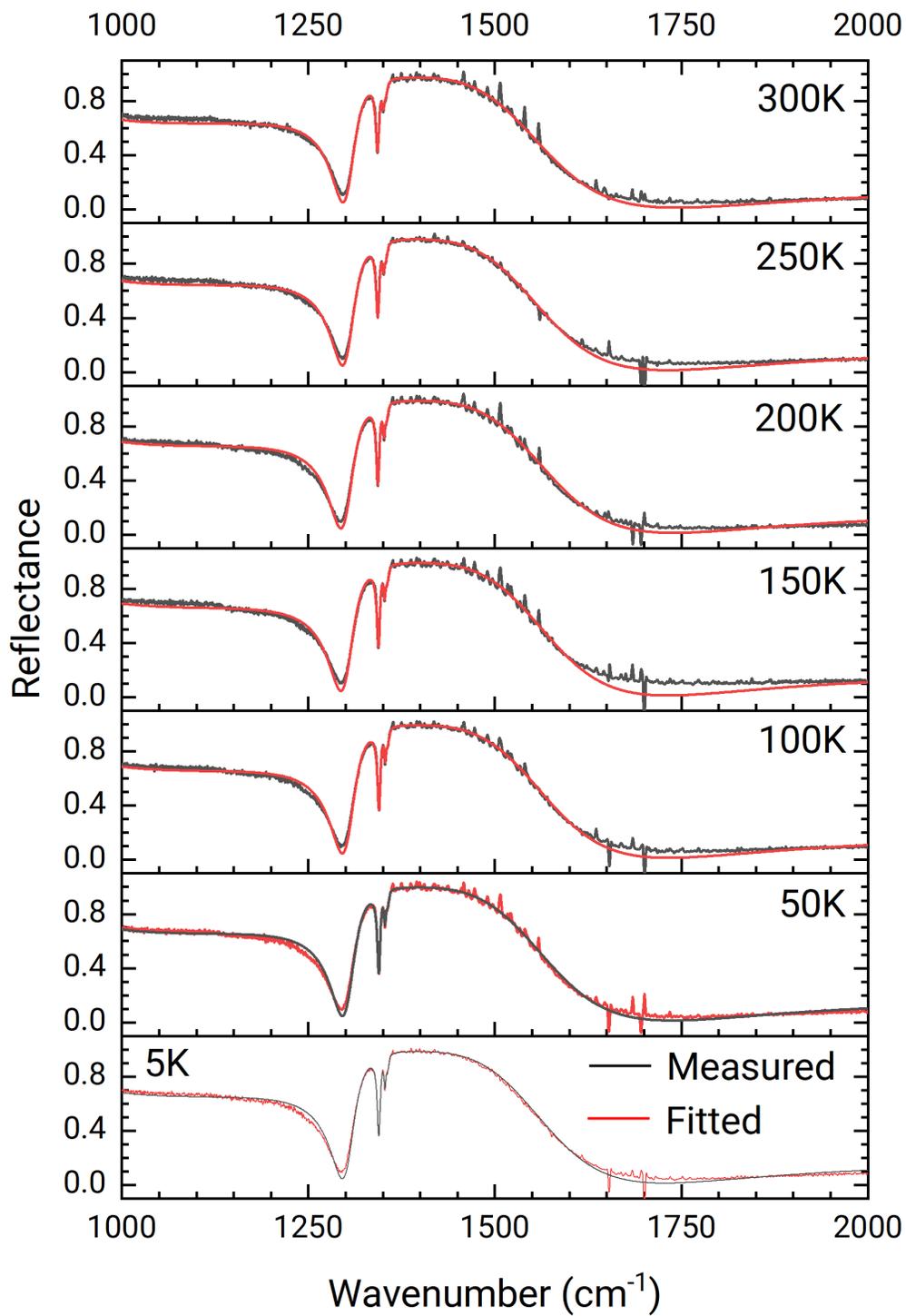

*Figure S6. Temperature-dependent spectra and fit data for the 750nm h$^{11}$BN flake*

**Supplemental Section 4 – High-frequency permittivity and LO phonon data**

As discussed in the main text, our fits are less consistent in the permittivity and LO phonon data. Fitted values are shown in Figure S6. The reason for this lack of sensitivity is that the LO phonon energy is most accurately determined by the zero-crossing of the dielectric function, which gives a near-zero reflectance from a bulk sample. The thin flakes do not completely absorb incident light, and the back reflections obscure this zero. This is not completely corrected through our data correction process. Further, this gives some uncertainty to the value of the high-frequency dielectric constant. However, we include the spectral range in the fit below the TO, where the dielectric function will approach $\varepsilon_{DC}$ – the low-frequency permittivity. The value of $\varepsilon_{DC}$ is given by the Lyddane-Sachs-Teller relations as;

$$\varepsilon = \varepsilon_\infty \frac{\omega_{LO}^2}{\omega_{TO}^2}$$

This shows that $\omega_{LO}$ and $\varepsilon_\infty$ values are coupled in the fit, explaining some systematic uncertainties in these parameters – particularly between the two thicknesses of h[10]BN flakes. The values we fit for the static dielectric function, also shown in **Figure S6**, show good consistency with prior work. This supports the idea that while errors exist in the fit, they are sufficiently accurate for predictive analysis of the nanophotonic properties. In particular, the results suggest that 750nm or larger flakes can give improved dielectric function estimates.

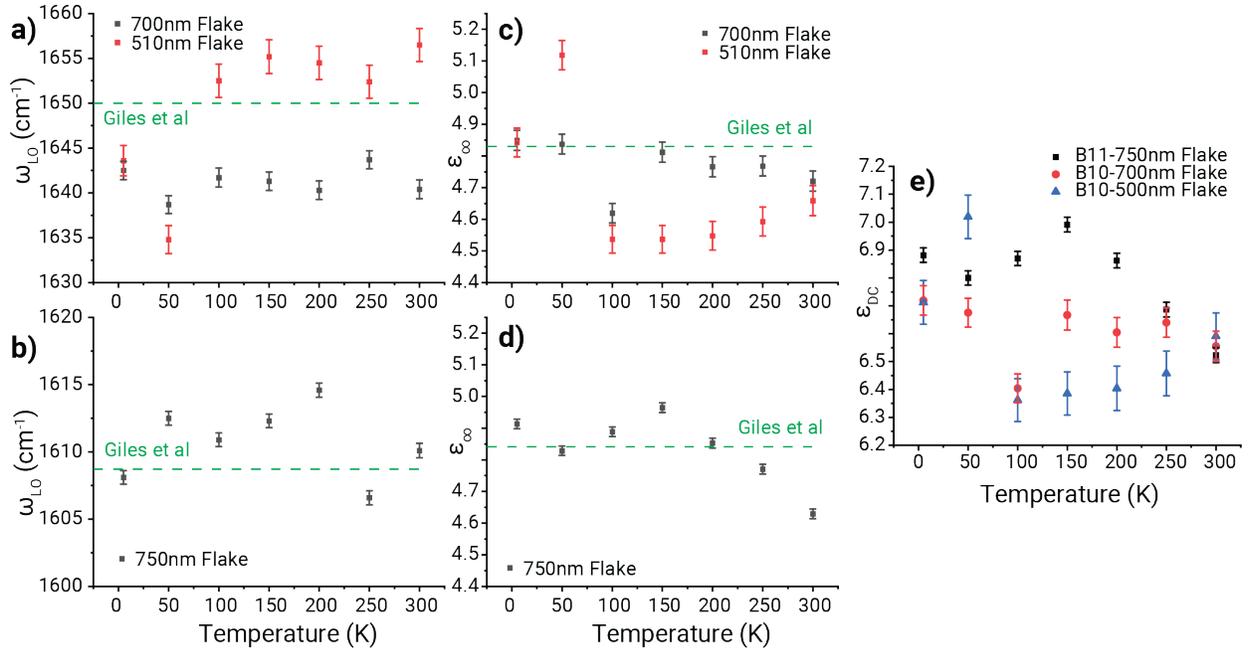

*Figure S6 - $\omega_{LO}$ (a/b), $\varepsilon_\infty$ (c/d), and $\varepsilon_{DC}$ (e) values for isotopically enriched $B^{10}$ and $B^{11}$ flakes respectively*

**Supplemental Section 5 – Validation via reflectance of flakes on Au substrate.**

In this section, we show that our fitted dielectric function can be used to adequately describe the optical properties of hBN, as shown in Figure S7. We can correctly predict the frequencies and linewidths of the infrared-measured FP modes, as well as the associated temperature tuning.

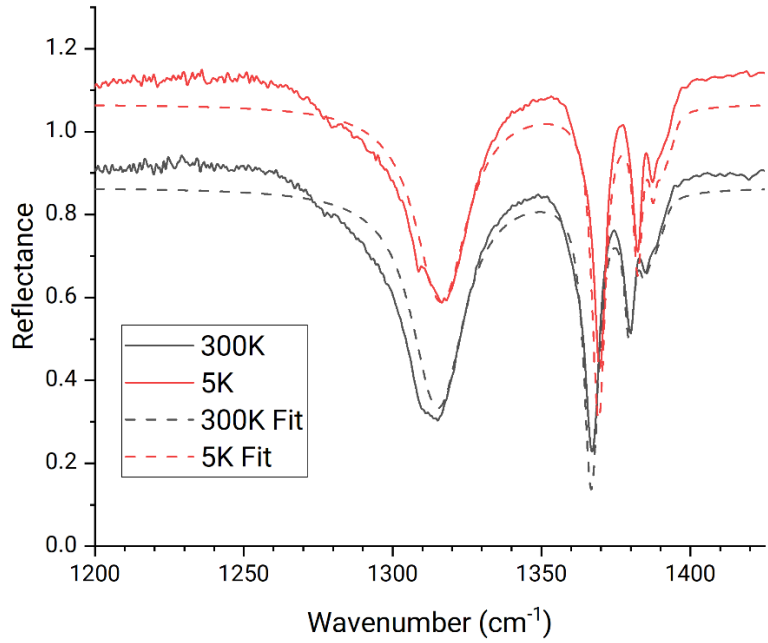

*Figure S7 – Validation of dielectric function fit using 1220nm thick B10 hBN data.*